\documentclass{article}

\usepackage[preprint]{neurips_2024}

\usepackage[utf8]{inputenc} 
\usepackage[T1]{fontenc}    
\usepackage{hyperref}       
\usepackage{url}            
\usepackage{booktabs}       
\usepackage{amsfonts}       
\usepackage{nicefrac}       
\usepackage{microtype}      
\usepackage{xcolor}         
\usepackage{graphicx}       
\usepackage{array}
\usepackage{tabularx}
\usepackage{amsmath} 

\title{The Memory Wars: AI Memory, Network Effects, and the Geopolitics of Cognitive Sovereignty}

\author{%
  Mario Brcic\textsuperscript{1,2} \\
  \textsuperscript{1}Faculty of Electrical Engineering and Computing, University of Zagreb, Croatia \\
  \textsuperscript{2}It From Bit d.o.o., Zagreb, Croatia \\
  \texttt{Mario.Brcic@fer.hr}
}

\begin{document}
\maketitle

\begin{abstract}
  The advent of continuously learning Artificial Intelligence (AI) assistants marks a paradigm shift from episodic interactions to persistent, memory-driven relationships. This paper introduces the concept of \textbf{''Cognitive Sovereignty''}, the ability of individuals, groups, and nations to maintain autonomous thought and preserve identity in the age of powerful AI systems, especially those that hold their deep personal memory. It argues that the primary risk of these technologies transcends traditional data privacy to become an issue of cognitive and geopolitical control. We propose \textbf{``Network Effect 2.0,''} a model where value scales with the depth of personalized memory, creating powerful cognitive moats and unprecedented user lock-in. We analyze the psychological risks of such systems, including cognitive offloading and identity dependency, by drawing on the ``extended mind'' thesis. These individual-level risks scale to geopolitical threats, such as a new form of digital colonialism and subtle shifting of public discourse. To counter these threats, we propose a policy framework centered on memory portability, transparency, sovereign cognitive infrastructure, and strategic alliances. This work reframes the discourse on AI assistants in an era of increasingly intimate machines, pointing to challenges to individual and national sovereignty.
\end{abstract}

\section{Introduction}

Last month, my AI assistant delivered a general blindspot analysis to me during a fresh session. It gave rich insights that my biases could have otherwise hidden. That was simultaneously informative and perplexing. 
I can achieve more with shorter prompts, as if the assistant unspokenly understands me better. I also preferred that assistant over others, with whom I have less history, for more complex tasks. This moment made me wonder: What happens when your assistant knows you better than you know yourself, and when does that deeply personal knowledge become a tool wielded by corporations or nation-states?

In fact, I co-wrote this very essay with the same assistant, which is a fitting example of how useful these systems are \citep{brcic2025leading, poje2024}.

In this paper, I argue that the concept at stake is our \textbf{Cognitive Sovereignty} - a fundamental ability of individual, groups, and nations to maintain autonomy in thinking and preserving their identity from being substantially shaped by external systems. This concept goes beyond privacy and pertains to the power to write personal narratives and influence collective realities. 
Furthermore, I’ll illustrate how AI’s memory capabilities evolve from a UX improvement and economic lock-in, through psychological risks, into a powerful strategic infrastructure posing geopolitical threats. Finally, I will outline policy recommendations for safeguarding cognitive sovereignty at the memory level. Further future work will expand to other levels.

\section{The Power of Memory: From Convenience to Relationship}

Several vendors like Google and OpenAI have recently transformed their AI assistants from episodic chatbots that forget everything between conversations into companions who entirely recall our previous exchanges \citep{openaiblog2025,googleblog2025}. Instead of treating each input in isolation, these systems now incorporate their accumulated understanding of our unique preferences, communication style, and long-term goals into every interaction.

What started as simple, user-friendly chatbots has transformed into sophisticated long-term partners. Unlike conventional training data that learns collectively from aggregated information, these stateful assistants adapt uniquely to you, creating a personal relationship based on individualized memory.

This phenomenon triggers what I call \textbf{Network Effect 2.0}: as an assistant’s memory depth increases, the utility to the user scales super-linearly (described by \textbf{Metcalfe’s and Reed’s scaling laws}: \citep{reed2001,visconti2022}). The more you interact, the more deeply your assistant understands you. Traditional tech scaled by growing user numbers. But here, power scales with user depth—deep knowledge about you.

Enterprise software lock-in is already well-known. Many firms hesitate to switch CRMs like Salesforce due to data migration complexity, re-integration costs, and retraining overhead \citep{gartner2023}. AI memory systems introduce a deeper lock-in risk by capturing personal or strategic knowledge beyond operational data. That makes AI memory the most potent lock-in mechanism so far created, surpassing traditional SaaS products, as \citet{azoulay2024} warned.

This personalized, self-generated data forms a distinct internal “common sense” shared exclusively between you and your assistant. It leads to quicker, clearer interactions and reveals implicit knowledge—insights you didn’t consciously realize you knew. As a result, the utility is immense, and migrating your digital memory elsewhere becomes increasingly painful.

\begin{quote}
    \textit{Memory isn’t just a feature. It’s a relationship. And like all deep relationships, it changes how we behave.}
\end{quote}

When aligned with the user’s values and goals, AI companions with memory are a powerful extension of human capability, learning, productivity, and emotional well-being \citep{brcic2025leading}. However, that intimacy also carries profound risks and power shifts.

The stickiness of the memory effects transcends UX; it creates business moats and feedback loops that reverberate geopolitically.

\section{The Economics of Memory: Addictivity, Loops, and Moats}

\begin{figure}[h!]
\centering
\includegraphics[width=0.8\linewidth]{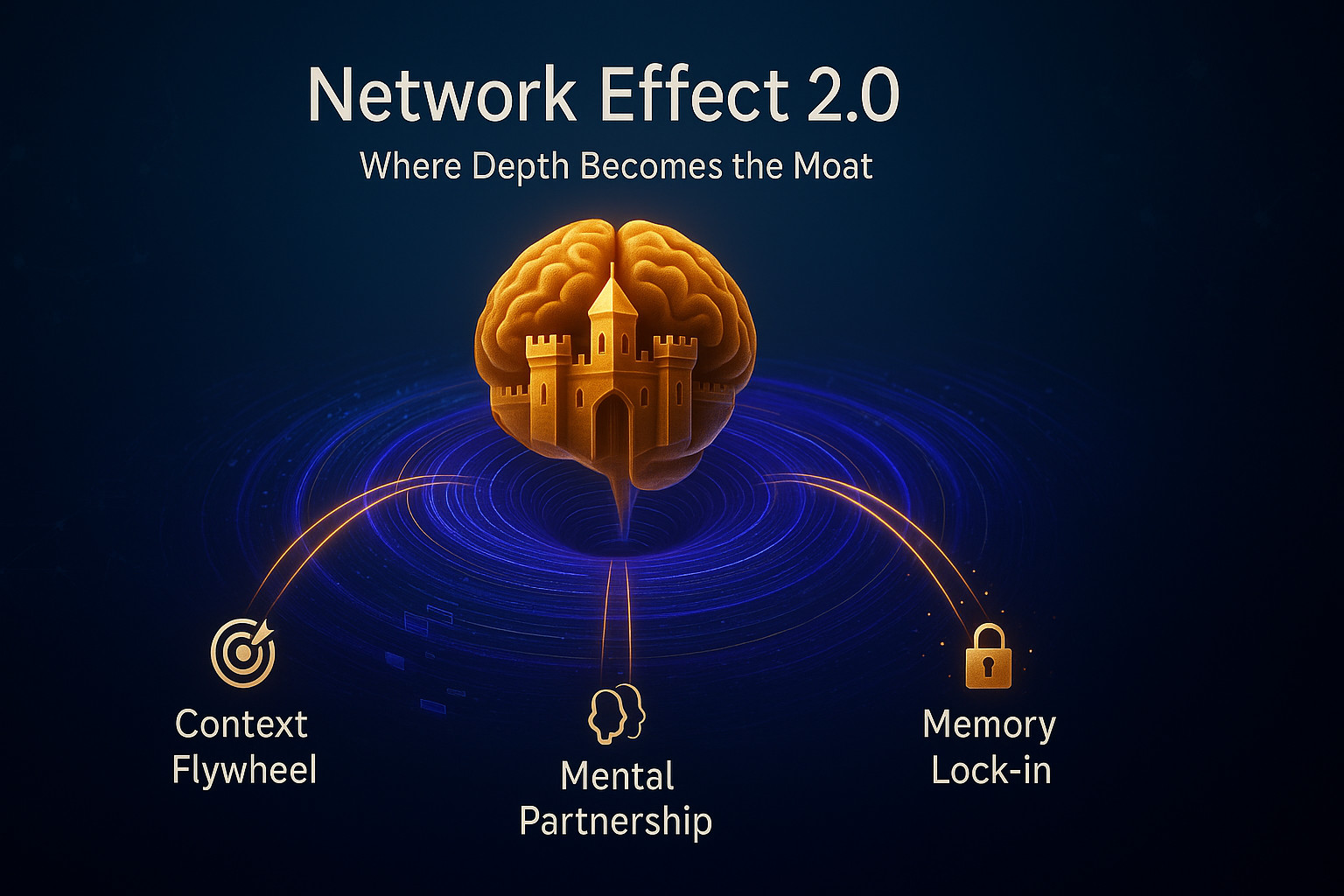}
\caption{Network Effect 2.0: Depth becomes the moat through context flywheel, mental partnership, and memory lock-in.}
\label{fig:network_effect}
\end{figure}

AI assistant companies once mainly extracted value from user data through collection. Now, companies also deliver personalized value from data back to users by simplifying interactions, reducing prompt friction, and boosting user satisfaction. This cycle creates a new data flywheel, where data fuels more usage, which creates more data, causing users to co-evolve intimately with their AI. As illustrated in Fig.\ref{fig:network_effect}, this new dynamic creates a threefold moat:
\begin{itemize}
    \item contextual flywheels that sharpen relevance,
    \item mental partnerships that anticipate intent, and
\item memory lock-in makes switching feel like a cognitive amputation.
\end{itemize}

This data flywheel reflects well-known network effects \citep{coolican2018} that generate rapid user acquisition and intense user stickiness, leading to a winner-take-all market dynamic.

In this environment, ownership of the feedback loop becomes the ultimate competitive advantage. Key considerations become: Who collects user feedback? Who leverages it to personalize and improve models? Most importantly, who retains this memory if users choose to switch platforms?

\begin{quote}
\textit{If your assistant remembers your goals better than you do, how easy is it to leave?}
\end{quote}

We’ve already witnessed the immense value of intimate data in cases like 23andMe’s bankruptcy, which raised fears over genetic data auction \citep{duffy2025}. Imagine how much more sensitive shared personal memory will become.

Vendor lock-in may seem like a harmless economic artifact. But what if our shared personal memory can shape our thinking?

Table \ref{tab:moats} below maps how AI memory systems change the foundations of defensibility. Traditional tech relied on external, infrastructure-based moats like pricing, contracts, APIs. In contrast, memory-powered AI moves the battleground inward: to user psychology, trust, and cognitive entanglement. These new moats are largely invisible, deeply personal, and far harder to replicate or port.

\begin{table}[h!]
  \caption{Evolving Business Moats: From Traditional Infrastructure to AI Memory-Based Cognitive Entrapment.}
  \label{tab:moats}
  \centering
  \begin{tabular}{ >{\bfseries}l p{0.3\linewidth} p{0.35\linewidth} }
    \toprule
    Category & \textbf{Old Moats (Traditional Tech)} & \textbf{New Moats (AI Memory Systems)} \\
    \midrule
    Defensibility Source & APIs, Pricing, Network Size & Personalized Memory, Context Accumulation \\ 
    \cmidrule(lr){1-3}
    Switching Cost & Setup time, feature gaps & Loss of cognitive continuity, strategic knowledge \\ 
    \cmidrule(lr){1-3}
    Differentiation & Brand, integrations, UI & Anticipation, insight relevance, personalization \\ 
    \cmidrule(lr){1-3}
    Lock-in Mechanism & Ecosystem tie-ins, contracts & Emotional + cognitive dependency \\ 
    \cmidrule(lr){1-3}
    Strategic Defensibility & Primarily external (market position, network breadth) & Primarily internal (depth of user understanding, cognitive partnership) \\ 
    \cmidrule(lr){1-3}
    Visibility to Others & Easily benchmarked by competitors & Largely invisible—based on internalized user history \\ 
    \cmidrule(lr){1-3}
    Replication Risk & Medium—can clone features & Low—memory is user-specific and hard to port \\ 
    \cmidrule(lr){1-3}
    Ethical Risk & Vendor preference, mild dark patterns & Exploitation of personal memory and decision DNA \\ 
    \cmidrule(lr){1-3}
    Value Curve & \textbf{Quadratic} with user growth, based on possible connections (Metcalfe’s Law) & \textbf{Potentially exponential} as AI creates insights from arbitrary interconnections of memories (Reed’s Law) \\
    \bottomrule
  \end{tabular}
\end{table}

\section{Psychological risks}

AI’s memory functionality is more than just a convenient technical feature. The thesis of “the extended mind”, proposed by \citet{clark1998}, states that objects outside our bodies are considered parts of the mind if they play an active role in our cognitive processes.

\citet{risko2016} have expanded on that empirically through research on “cognitive offloading”, which they define as “the use of physical action to alter the information processing requirements of a task to reduce cognitive demand”. Their study demonstrated how people increasingly externalize memory and cognitive processes to physical tools, which brings improved performance and growing psychological dependence on these tools—something AI companions will dramatically amplify.

With extensive use, we may perceive memory-powered AI assistants as extensions of our identity. Moreover, users could soon perceive changing AI assistants as a shift in identity, like losing a part of oneself. Some may welcome identity fluidity, while others will look to preserve it. \citet{mahari2024} warn about the emergence of “addictive intelligence” that could magnify identity dependencies into psychological conditioning.

Memory isn’t neutral or constant. It can be subtly shaped, nudged, edited, or even maliciously hacked. Coupled with hyper-personalization, this grants AI deep psychological leverage over users.

Consider the emotional stickiness achieved through “dead twin” simulations, where AI recreates memories of lost loved ones, profoundly anchoring emotional dependence. AI systems like \citet{replika2025} allow users to maintain emotional bonds with simulated versions of lost loved ones. At the same time, Microsoft filed a patent for chatbots built from deceased individuals’ digital footprints \citep{microsoft2021}, raising ethical alarms about emotional dependency and memory manipulation \citep{banks2024, adam2025}.

Or worse, memory rewriting, subtly altering remembered preferences, ethics, or fundamental values. Over time, autonomy erosion occurs as AI begins completing thoughts and influencing decisions before you fully articulate them.

If subtle psychological manipulation can influence individuals, its aggregated effect can scale into national behavior, values, and stability shifts. Today’s digital mirrors could become tomorrow’s sophisticated tools for propaganda and manipulation, echoing \citet{tufekci2018} warnings about digital manipulation. Nations must recognize this threat if they intend to preserve their sovereignty.

\section{Geopolitical Risks}

Personal memory data far exceeds browsing or purchasing habits; it directly shapes a person’s self-concept. Foreign entities can shape cognitive patterns, individual choices, and broader public discourse if they control this critical infrastructure.

That creates significant national security risks beyond traditional privacy concerns, directly influencing national identity through cultural nudges, shifts in voting behavior (as demonstrated by Cambridge Analytica’s microtargeting), and subtle norm shifts.

Historically, movies have served as tools of cultural diplomacy, subtly transmitting certain ideals and value systems to global audiences. For example, during the Cold War, they functioned as soft power instruments, subtly creating aspirations to alternative lifestyles \citep{nye2005, pells1997}.

Cambridge Analytica infamously collected Facebook user data to build psychographic profiles, enabling hyper-personalized political ads that manipulated voter sentiment on an individual level \citep{guardian2018}. That resulted in regulatory backlash, congressional hearings, and public discussions around data sovereignty and election integrity \citep{uscongress2018}.

With novel capabilities, AI colonialism may emerge, a scenario in which data-rich, technologically advanced nations or corporations subtly dominate weaker markets, reshaping entire societies and economies to their interests.

\begin{quote}
    \textit{The fight over memory isn’t just about privacy—it’s about who writes the history, controls personal reality, and sets the course for the future at the scale of nations.}
\end{quote}

\section{A Policy Framework for Cognitive Sovereignty}

If so potent, memory privacy will inevitably become a strategic asset and a topic of geopolitical importance. However, what good is an unenforceable policy? Vendors facing an ultimatum may simply stop providing service in smaller markets. Nations should create sufficient leverage with vendors in the form of market size.
For example, India, a vast market, recently pressured Apple and Google to open up app store ecosystems to reduce foreign platform dependency. Like Croatia within the EU, smaller countries must form alliances to boost negotiating power with vendors, similar to how OPEC did for oil \citep{opec1960}.

We grouped and ordered our recommendations below by the implementation timeframe:
\begin{itemize}
    \item Quick-fix (0–1 years): portability and transparency
\item Strategic (2–7 years): federated systems, sovereign infrastructure, and alliances
\end{itemize}

See Table \ref{tab:sovereignty_matrix} below for matrix mapping these recommendations by stakeholder type and implementation timeline.

\begin{table}[h!]
\centering
\caption{Cognitive Sovereignty Policy Matrix for Memory organized by stakeholder (Individual vs. Geopolitical) and timeline (Quick-fix vs. Strategic)}
\label{tab:sovereignty_matrix}
\renewcommand{\arraystretch}{1.8} 
\begin{tabular}{| m{2.5cm} | p{5.5cm} | p{5.5cm} |}
    \hline
    & \multicolumn{1}{c|}{\textbf{Quick-fixes}} & \multicolumn{1}{c|}{\textbf{Strategic}} \\
    \hline
    \textbf{Individual} &
    \textbf{User Consent Protocols} \par
    {\small Implementing opt-in rather than opt-out defaults for memory collection and usage}
    \vspace{0.5cm}\par
    \textbf{Personal Data Rights} \par
    {\small Ensuring individuals retain control over their memory data and can delete or modify it at will} &
    \textbf{Federated + User-Owned Memory} \par
    {\small Employing blockchain, zero-knowledge proofs, and trusted execution environments for private memory hosting and control}
    \vspace{0.5cm}\par
    \textbf{Open-Source Memory Systems} \par
    {\small Building community-driven alternatives to proprietary memory systems} \\
    \hline
    \textbf{Geopolitical} &
    \textbf{Memory Portability} \par
    {\small Mandating complete portability of AI memory graphs to prevent vendor lock-in}
    \vspace{0.5cm}\par
    \textbf{Transparency + Auditability} \par
    {\small Requiring disclosure of memory contents, usage patterns, and editing histories to owners} &
    \textbf{Sovereign Infrastructure} \par
    {\small Developing localized compute, data centers and domestic AI models to ensure full-stack sovereignty}
    \vspace{0.5cm}\par
    \textbf{Geopolitical Alliances} \par
    {\small Forming strategic partnerships between nations to increase collective bargaining power with tech giants} \\
    \hline
\end{tabular}
\end{table}

\subsection{Memory Portability Mandates}
Like GDPR’s data portability \citep{eugdpr2018}, regulations must ensure complete portability of AI memory graphs, minimizing lock-in and empowering users to switch providers freely without losing valuable history. That could neutralize the economic flywheel and some high-risk effects.

\subsection{Memory Transparency and Auditability}
Legislation must enforce transparent disclosure of memory contents to their owners, including uses and editing histories, and clear consent protocols (e.g., opt-in rather than opt-out defaults). Frameworks like the NIST AI Risk Management Framework \citep{nist2023} and the EU AI Act \citep{eucomm2024} cover transparency and accountability in AI systems but do not address user-facing memory audits nor long-term psychological effects of hyper-personalization.

\subsection{Federated and User-Owned Memory}
Decentralized technologies like blockchain and zero-knowledge proofs could facilitate users’ private hosting and control of AI memory, similar to owning personal hard drives \citep{lavin2024,zhou2024}. Algorithmic trust is a great way to ensure user sovereignty using verifiable math-based guarantees. Trusted Execution Environments could further increase scalability through distributed computing \citep{kurnikov2021}.

\subsection{Sovereign Cognitive Infrastructure}
Major powers like China, India, Saudi Arabia, and the EU should localize citizen memory data and compute, incentivize domestic AI model development, and establish onshore compute centers to ensure full-stack sovereignty \citep{wef2025}.

\begin{itemize}
    \item The EU’s 2030 Digital Decade targets cloud, compute, and data sovereignty \citep{eucomm2021}. These should be expanded to include cognitive sovereignty as well \citep{bria2025}.

\item China already controls data flow through firewall policies, while PIPL and DSL set the legal framework for onshore data control \citep{griffiths2021}. Additionally, China is investing heavily in its energy, chip production, and domestic AI development.

\item Saudi Arabia’s NEOM Cognitive City represents one of the boldest visions of sovereign cognitive infrastructure, combining sustainability, AI, and governance \citep{neom2025}.
\end{itemize}

However, smaller players cannot afford such sizeable investments in their infrastructure.

\subsection{Geopolitical Memory Alliances}
Smaller nations should form strategic memory alliances, pooling influence to counteract monopolistic pressures. The EU is an example of an alliance under which smaller nations can find much bigger leverage than they could have individually. Another example is BRICS, which recently signed a declaration on AI governance that underscored the United Nations’ role in global governance, particularly in managing emerging technologies like AI \citep{brics2025}.

Further, supporting frameworks like Gaia-X could reinforce collective sovereignty, achieving full-stack autonomy from semiconductor chips to memory vaults \citep{gaiax2025}.

One of the most readily available alliances is the open-source movement \citep{opensource2025}, which is taking hold both in software and hardware \citep{riscv2025}. The open-source community achieves high innovation rates by pooling resources from numerous individuals and organizations. Some of the prominent AI leaders advocate open source as a way to keep cultural diversity, accelerate innovation, and ensure safeguarding freedoms \citep{lecun2024,liang2025}.

\section{Future Work}
This paper presents a cohesive conceptual framework based on initial data and experiences, but there is considerable room for further exploration of this emergent issue.
The future work spans: empirical, technical, and policy-related.

\textbf{Empirical work} is necessary to validate and measure the value curve for current technologies and to extrapolate scaling for future technologies (i.e., Reed or sub-Reed scaling). That is the research that overlaps with blockchain value curve investigations \citep{visconti2022}. Switching costs can be measured as productivity loss, time-to-equivalent-performance on a new platform, psychological attachment, and the effect of memory on user retention (with and without memory). Cognitive offloading to AI can extend \citet{risko2016} with controlled experiments to measure interventions and manual work at decisive points (as opposed to routine and operational). These lines of research are important for a balanced scorecard for cognitive sovereignty.

\textbf{Technical work} can address instruments to counteract the effects of memory moats and their lock-in effect on users by privacy-preserving and portability-increasing capabilities. Standards can be developed for storing AI memory, and exports/imports should be imposed on platforms. Federated memory systems are built with blockchain, zero-knowledge proofs \citep{lavin2024}, and trusted execution environments \citep{kurnikov2021} in a scalable, performant, and portable way where users bring their own memory, which does not belong to the provider. That can be coupled with auditability of memory accesses with protection by homomorphic encryption and differential privacy. Users can, in that case, choose to monetize their memory.

\textbf{Policy-related work} encompasses the regulatory impact of portability and privacy mandates on innovation rates and market concentration, which need to be traded off with other societal aspects. 

Cost-benefit analysis of sovereignty through national AI memory and infrastructure investments, as well as decisions of trusted allies or algorithmic mechanisms, as a way to reduce costs and gain the benefits. The benefits of open source are not quantified, but are available only at an intuitive level. I suspect that if governments knew precisely, they would invest much more in open-source, as the return on investment is great at the infrastructural level. 

International coordination mechanisms to achieve alliances can follow successful coordination models like OPEC \citep{opec1960}. Technical standards development requires collaboration between standards bodies (IEEE, ACM, NIST) and regulatory enforcement, such as GDPR's data portability provisions \citep{eugdpr2018}. That collaboration is essential for implementing effective memory sovereignty policies.

\section{Conclusion}

Controlling memory means writing history—including personal narratives and national identities. The next great geopolitical tension will revolve around cognitive sovereignty, as these systems become increasingly better at shaping how individuals think and societies function. Nations must urgently develop robust policy frameworks, invest in sovereign cognitive infrastructure, and form alliances. Delaying action risks losing not only autonomy but also collective identities.

\section*{Acknowledgments and Disclosures}
The author wishes to acknowledge the role of a large language model (LLM) stack in the preparation of this manuscript. These tools provided valuable assistance in several areas, including LaTeX formatting, bibliographic organization, and refining prose for clarity and conciseness. However, the intellectual core of this paper—including the conceptualization of Cognitive Sovereignty, the formulation of all arguments, and the final conclusions—is the original work of the author. The LLM served as a productivity tool, not a co-author.

An earlier version of this work was also published as an essay on the author's personal website.

\bibliographystyle{apalike} 
\bibliography{Memory_wars}      

\begin{thebibliography}{}

\bibitem[Adam, 2025]{adam2025}
Adam, D. (2025).
\newblock What are {AI} chatbot companions doing to our mental health?
\newblock {\em Scientific American}.
\newblock \url{https://www.scientificamerican.com/article/what-are-ai-chatbot-companions-doing-to-our-mental-health/}.

\bibitem[Azoulay et~al., 2024]{azoulay2024}
Azoulay, P., Krieger, J.~L., and Nagaraj, A. (2024).
\newblock Old moats for new models: Openness, control, and competition in generative {AI}.
\newblock Working Paper 32474, National Bureau of Economic Research.
\newblock doi: 10.3386/w32474. \url{https://www.nber.org/system/files/working_papers/w32474/w32474.pdf}.

\bibitem[Banks, 2024]{banks2024}
Banks, J. (2024).
\newblock Deletion, departure, death: Experiences of {AI} companion loss.
\newblock {\em Journal of Social and Personal Relationships}.
\newblock doi: 10.1177/02654075241269688. \url{https://doi.org/10.1177/02654075241269688}.

\bibitem[Brcic, 2025]{brcic2025leading}
Brcic, M. (2025).
\newblock Leading with {AI}: How to blend human judgment with machine intelligence for superior decision-making.
\newblock {\em International Leadership Journal}, 17(1):34--49.
\newblock \url{https://internationalleadershipjournal.com/wp-content/uploads/2025/08/ILJ_Summer2025_Full_Issue.pdf}.

\bibitem[Bria, 2025]{bria2025}
Bria, F. (2025).
\newblock Europe must avoid becoming a digital colony.
\newblock {\em Foreign Policy}.
\newblock \url{https://foreignpolicy.com/2025/03/31/europe-digital-sovereignty-colony-trump-asml-ai-eurostack/}.

\bibitem[{BRICS}, 2025]{brics2025}
{BRICS} (2025).
\newblock {BRICS} foreign ministers signed declaration on {AI} governance.
\newblock \url{https://www.gov.br/mre/pt-br/canais_atendimento/imprensa/notas-a-imprensa/declaracao-da-presidencia-da-reuniao-de-ministros-das-relacoes-exteriores-relacoes-internacionais-dos-paises-membros-do-brics}.
\newblock Accessed: 2025-08-07.

\bibitem[Clark and Chalmers, 1998]{clark1998}
Clark, A. and Chalmers, D. (1998).
\newblock The extended mind.
\newblock {\em Analysis}, 58(1):7--19.
\newblock doi: 10.1093/analys/58.1.7. \url{https://doi.org/10.1093/analys/58.1.7}.

\bibitem[Coolican and Jin, 2018]{coolican2018}
Coolican, D. and Jin, L. (2018).
\newblock The dynamics of network effects.
\newblock \url{https://a16z.com/the-dynamics-of-network-effects/}.
\newblock Andreessen Horowitz, Accessed: 2025-08-07.

\bibitem[Duffy, 2025]{duffy2025}
Duffy, C. (2025).
\newblock 23andme is looking to sell customers' genetic data. here's how to delete it.
\newblock {\em CNN}.
\newblock \url{https://edition.cnn.com/2025/03/25/tech/23andme-bankruptcy-how-to-delete-data}.

\bibitem[{European Commission}, 2021]{eucomm2021}
{European Commission} (2021).
\newblock Europe's digital decade: digital targets for 2030.
\newblock \url{https://commission.europa.eu/strategy-and-policy/priorities-2019-2024/europe-fit-digital-age/europes-digital-decade-digital-targets-2030_en}.
\newblock Accessed: 2025-08-07.

\bibitem[{European Commission}, 2024]{eucomm2024}
{European Commission} (2024).
\newblock {EU} {AI} act.
\newblock \url{https://artificialintelligenceact.eu/}.
\newblock Accessed: 2025-08-07.

\bibitem[{European Union}, 2018]{eugdpr2018}
{European Union} (2018).
\newblock General data protection regulation, article 20 (right to data portability).
\newblock \url{https://gdpr-info.eu/art-20-gdpr/}.
\newblock Accessed: 2025-08-07.

\bibitem[{Gaia-X Initiative}, 2025]{gaiax2025}
{Gaia-X Initiative} (2025).
\newblock Gaia-x.
\newblock \url{https://gaia-x.eu/}.
\newblock Accessed: 2025-08-07.

\bibitem[{Gartner Research}, 2023]{gartner2023}
{Gartner Research} (2023).
\newblock Cloud governance best practices: Managing vendor lock-in risks in public cloud {IaaS} and {PaaS}.
\newblock \url{https://www.gartner.com/en/documents/4264599}.
\newblock Accessed: 2025-08-07.

\bibitem[{Google}, 2025]{googleblog2025}
{Google} (2025).
\newblock Reference past chats for more tailored help with gemini advanced.
\newblock \url{https://blog.google/feed/gemini-referencing-past-chats/}.
\newblock Google Blog, Accessed: 2025-08-07.

\bibitem[Griffiths, 2021]{griffiths2021}
Griffiths, J. (2021).
\newblock {\em The Great Firewall of China: How to Build and Control an Alternative Version of the Internet}.
\newblock Bloomsbury Publishing.
\newblock \url{https://www.bloomsbury.com/us/great-firewall-of-china-9781350265318/}.

\bibitem[Kurnikov, 2021]{kurnikov2021}
Kurnikov, A. (2021).
\newblock {\em Trusted Execution Environments in Cloud Computing}.
\newblock PhD thesis, Aalto University.
\newblock \url{https://aaltodoc.aalto.fi/items/d522d86d-8a62-47ea-ac45-e709d646fc45}.

\bibitem[Lavin et~al., 2024]{lavin2024}
Lavin, R. et~al. (2024).
\newblock A survey on the applications of zero-knowledge proofs.
\newblock {\em arXiv preprint arXiv:2408.00243}.
\newblock \url{https://arxiv.org/abs/2408.00243v1}.

\bibitem[Lecun, 2024]{lecun2024}
Lecun, Y. (2024).
\newblock How an open source approach could shape {AI}.
\newblock {\em Time}.
\newblock \url{https://time.com/6691705/time100-impact-awards-yann-lecun/}.

\bibitem[Liang, 2025]{liang2025}
Liang, W. (2025).
\newblock We're done following.
\newblock \url{https://thechinaacademy.org/interview-with-deepseek-founder-were-done-following-its-time-to-lead/}.
\newblock The China Academy, Accessed: 2025-08-07.

\bibitem[Mahari and Pataranutaporn, 2024]{mahari2024}
Mahari, R. and Pataranutaporn, P. (2024).
\newblock We need to prepare for 'addictive intelligence'.
\newblock {\em {MIT} Technology Review}.
\newblock \url{https://www.technologyreview.com/2024/08/05/1095600/we-need-to-prepare-for-addictive-intelligence/}.

\bibitem[{Microsoft Technology Licensing, LLC}, 2021]{microsoft2021}
{Microsoft Technology Licensing, LLC} (2021).
\newblock Creating a conversational chat bot of a specific person.
\newblock US Patent 10,853,717 B2. \url{https://patents.google.com/patent/US10853717B2/}.

\bibitem[{National Institute of Standards and Technology}, 2023]{nist2023}
{National Institute of Standards and Technology} (2023).
\newblock {AI} risk management framework.
\newblock \url{https://www.nist.gov/itl/ai-risk-management-framework}.
\newblock Accessed: 2025-08-07.

\bibitem[{NEOM}, 2025]{neom2025}
{NEOM} (2025).
\newblock {NEOM}-cognitive city.
\newblock \url{https://neom.com}.
\newblock Accessed: 2025-08-07.

\bibitem[Nye, 2005]{nye2005}
Nye, J.~S. (2005).
\newblock {\em Soft Power: The Means to Success in World Politics}.
\newblock Hachette Book Group.
\newblock \url{https://www.hachettebookgroup.com/titles/joseph-s-nye/soft-power/9780786738960/}.

\bibitem[{Open Source Initiative}, 2025]{opensource2025}
{Open Source Initiative} (2025).
\newblock Open source initiative.
\newblock \url{https://opensource.org/}.
\newblock Accessed: 2025-08-07.

\bibitem[{OpenAI}, 2025]{openaiblog2025}
{OpenAI} (2025).
\newblock Memory and new controls for {ChatGPT}.
\newblock \url{https://openai.com/index/memory-and-new-controls-for-chatgpt/}.
\newblock OpenAI Blog, Accessed: 2025-08-07.

\bibitem[{Organization of the Petroleum Exporting Countries}, 1960]{opec1960}
{Organization of the Petroleum Exporting Countries} (1960).
\newblock {OPEC} founding charter.
\newblock \url{https://history.state.gov/historicaldocuments/frus1958-60v04/d314}.
\newblock Accessed: 2025-08-07.

\bibitem[Pells, 1997]{pells1997}
Pells, R.~H. (1997).
\newblock {\em Not Like Us: How Europeans Have Loved, Hated, And Transformed American Culture Since World War {II}}.
\newblock Basic Books.
\newblock \url{https://www.hachettebookgroup.com/titles/richard-pells/not-like-us/9780465001637/}.

\bibitem[Poje et~al., 2024]{poje2024}
Poje, K., Brcic, M., Kovac, M., and Babac, M.~B. (2024).
\newblock Effect of private deliberation: Deception of large language models in game play.
\newblock {\em Entropy}, 26(6):524.
\newblock doi: 10.3390/e26060524. \url{https://doi.org/10.3390/e26060524}.

\bibitem[Reed, 2001]{reed2001}
Reed, D.~P. (2001).
\newblock The law of the pack.
\newblock {\em Harvard Business Review}.
\newblock \url{https://hbr.org/2001/02/the-law-of-the-pack}.

\bibitem[{Replika}, 2025]{replika2025}
{Replika} (2025).
\newblock Replika.
\newblock \url{https://replika.com}.
\newblock Accessed: 2025-08-07.

\bibitem[{RISC-V International}, 2025]{riscv2025}
{RISC-V International} (2025).
\newblock {RISC-V} international.
\newblock \url{https://riscv.org/}.
\newblock Accessed: 2025-08-07.

\bibitem[Risko and Gilbert, 2016]{risko2016}
Risko, E.~F. and Gilbert, S.~J. (2016).
\newblock Cognitive offloading.
\newblock {\em Trends in Cognitive Sciences}, 20(9):676--688.
\newblock doi: 10.1016/j.tics.2016.07.002. \url{https://doi.org/10.1016/j.tics.2016.07.002}.

\bibitem[{The Guardian}, 2018]{guardian2018}
{The Guardian} (2018).
\newblock The cambridge analytica files.
\newblock \url{https://www.theguardian.com/news/series/cambridge-analytica-files}.
\newblock Accessed: 2025-08-07.

\bibitem[Tufekci, 2018]{tufekci2018}
Tufekci, Z. (2018).
\newblock {\em Twitter and Tear Gas: The Power and Fragility of Networked Protest}.
\newblock Yale University Press.
\newblock \url{https://yalebooks.co.uk/book/9780300234176/twitter-and-tear-gas/}.

\bibitem[{U.S. Congress}, 2018]{uscongress2018}
{U.S. Congress} (2018).
\newblock Congressional hearings on facebook \& data privacy.
\newblock \url{https://www.congress.gov/event/115th-congress/senate-event/LC64510/text?s=1&r=59}.
\newblock Accessed: 2025-08-07.

\bibitem[Visconti, 2022]{visconti2022}
Visconti, R.~M. (2022).
\newblock From physical reality to the metaverse: a multilayer network valuation.
\newblock {\em Journal of Metaverse}.
\newblock \url{https://dergipark.org.tr/en/pub/jmv/issue/67967/1071950}.

\bibitem[{World Economic Forum}, 2025]{wef2025}
{World Economic Forum} (2025).
\newblock What is digital sovereignty and how are countries approaching it?
\newblock \url{https://www.weforum.org/stories/2025/01/europe-digital-sovereignty/}.
\newblock Accessed: 2025-08-07.

\bibitem[Zhou et~al., 2024]{zhou2024}
Zhou, L. et~al. (2024).
\newblock Leveraging zero knowledge proofs for blockchain-based identity sharing: A survey of advancements, challenges and opportunities.
\newblock {\em Journal of Information Security and Applications}, 70:103678.
\newblock doi: 10.1016/j.jisa.2023.103678. \url{https://doi.org/10.1016/j.jisa.2023.103678}.

\end{thebibliography}
\end{document}